# Design techniques for a seamless information system architecture

Grigory Tsiperman

Moscow, Russia

Assistant professor

National University of Science and Technology "MISIS"

gntsip@gmail.com

**Abstract.** The paper discusses design techniques for a seamless architecture of information systems (IS). A seamless architecture is understood as such an architectural description of an IS, that defines explicit connections between elements of architectural models of various architectural representations. Design techniques are based on the adaptive clustering method developed by the author, which allows one to bridge technological gaps between architectural abstracts of different levels and to link architectural models in such a way as to ensure the design of a more detailed model based on a model of a higher level of abstraction.

## 1. INTRODUCTION

The concept of IS architecture historically preceded the concept of enterprise architecture, which arose from the realization that the IS model must meet the requirements of the business and be able to flexibly adapt to its needs. This accordance should be based on a deep understanding of the business and the prospects for its development. IS should ensure the implementation of business requirements, have the ability to adapt to its changes.

The idea of enterprise architecture is to create interrelated architectural models that combine the concepts of mission, goals, business strategy of the enterprise, business processes, information systems, etc. To implement the idea of enterprise architecture, several methodologies have been proposed (GERAM [1], TOGAF [2], FEA [3]), none of which is devoid of shortcomings and is not a paradigm today. An overview of these methodologies is given in [4]. This paper discusses the issues of building an IS architecture as part of an enterprise architecture.

One way or another, starting with Zachman [5], the methodologies for creating an IS architecture assume a top-down approach, with the sequential construction of architectural models of an increasing level of detail. The practice of designing an IS architecture presupposes a heuristic transition between architectural models of various levels of abstraction: a more detailed architecture is built to ensure the closest correspondence of technical solutions to a model of the previous level. In doing so, the architect uses creativity, experience and knowledge to build a detailed model, and then proves (or considers it obvious) that the resulting model satisfies the requirements arising from a more abstract architecture.

Many authors ([6], [7], [8]) have posed the task of combining various technologies for constructing architectural representations of IS into a single technological stream, eliminating technique gaps between architectural models of various levels of abstraction. The technology for constructing a seamless IS architecture considered in this work is an attempt to solve this problem.

Theoretical foundations of a seamless architecture are considered in [9]. Seamless IS architecture is understood as an architectural description in which explicit relationships are defined between elements of architectural models of various architectural representations.

In the case of a seamless relation, description of the IS architecture is an interconnected set of architectural representations, and, for elements of architectural models, tracing becomes possible. This article discusses methods for designing functional IS architectures that have been tested on a large number of projects over the past few years. These methods are based on the use of the adaptive clustering method (ACM, see [10], [11]), in which detailed architectural models are built on the basis of inference from high-level abstraction models, providing seamless relations and traceability between the components of architectural models.

## 2. FORMULATION OF THE PROBLEM

The paper describes techniques for designing a seamless architecture and provides fragments of artifacts of the IS design process (**Fig. 1**).

IS design begin with business architecture modeling, which results in an architectural description of a business process, taking into account solutions for its automation. Business architecture serves as the basis for the development of the IS concept.

The second step is the gathering and formulation of user requirements for IS, which is based on a business architecture supplemented by operational services. Operational services define the automated functions of the business process that are confirmed with stakeholders.

At the third step, the external appearance of the IS is defined based on the design of the functional architecture. The functional architecture describes the interconnected IS dialogs that define the view functions.

The view functions are decomposed into IS software modules at the fourth design step. For decomposition of view functions, the component structure of the IS must be defined.

The fifth step is to design the methods of the data architecture classes. Software modules are decomposed into the structure of classes, defining the methods of these classes.

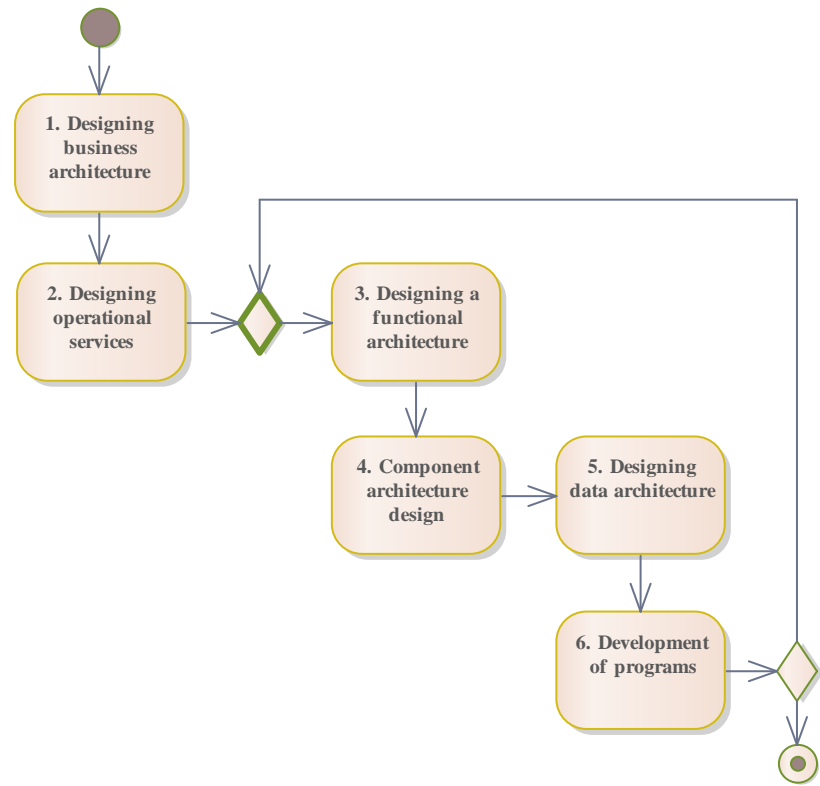

**Fig. 1.** Seamless architecture design process

Based on this component architecture and data architecture, the IS operational documentation is generated.

Further development of programs is carried out. The results of the development of programs may require serious changes in the developed architectural models, therefore, operational documents should be adjusted correspondingly.

Design technologies are considered using the example of a project for state registration of vehicles integrated into the portal of public services, i.e. contacting the service and all the information about residents and their vehicles necessary for registration is contained directly on the portal of public services.

The purpose of the state registration of vehicles is to ensure the implementation of the legislation that regulates relations arising in connection with the operation of vehicles. The need for registration arises during the initial acquisition of a new vehicle by a resident. Re-registration is required when changing the owner of the vehicle. Registration / re-registration of the vehicle is carried out on the basis of the provision by the owner of the vehicle of documents confirming ownership. As a result of registration / re-registration of the vehicle, the owner receives new registration marks and an entry of the new owner is made in the vehicle passport.

For registration, it is necessary to collect documents specified by law, pay a state duty, plan a visit to the traffic police to register at the appointed time. The registration process consists of checking the vehicle, entering the registered vehicle in the register and issuing to the resident the Vehicle Registration Certificate, Vehicle passport and registration marks.

The main scenario (the first level of decomposition) of the vehicle registration business process is shown in Fig. 2.

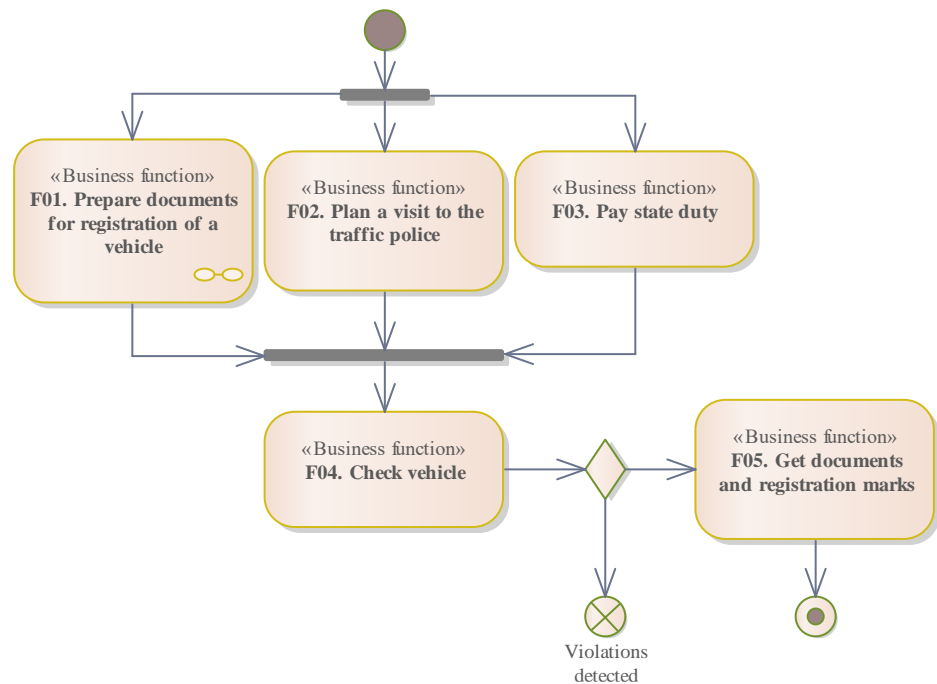

Fig. 2. Vehicle registration business process

The paper provides an example of seamless design of the business function "F01. Prepare documents for registration of a vehicle". The function is defined by the following comment:

To register a vehicle, you must prepare the following documents

− passport of a citizen;

− notarized power of attorney (for authorized representatives);

− vehicle passport;

− document proving ownership of the motor vehicle;

− insurance policy.

The necessary documents are collected using the public services portal from the resident's profile, where information about these documents must be posted and their authenticity must be verified.

If not, all documents are present in the profile of a citizen, she or he has the opportunity to submit documents directly to the traffic police.

## 3. BUSINESS DRIVEN ARCHITECTURE DESIGN

Let's consider the decomposition of the business function "F01. Prepare documents for registration of a vehicle", presented in **Fig. 3**. The decomposition includes the implemented business operations related to the portal of public services, and the projected operations of the vehicle registration service.

A business operation is the place where the interaction of the user with the information system occurs, and therefore the description of these operations determines this interaction. For example, the description of the business transaction "OPTC01.03. Initiate the vehicle registration process" can be represented by the following text:

The process of registration/re-registration of the vehicle begins with the verification of the initiating message: the message must be sent from the portal of public services. If authenticity of the message cannot be verified, the characteristics of the message and its source are recorded in the system log. The service stops working without any messages.

If the message verification is successful, the User is provided with all the necessary information about the processes of registration and re-registration of the vehicle and is prompted to proceed to the verification of the primary documents for the process, payment of the state fee and the choice of time to visit the traffic police.

Business architecture details the functional requirements for the IS, which are reflected in the functional architecture. The description of business operations allows us to highlight the automated functions that define the functional requirements for the system. The set of functions to be automated is defined in the operational service associated with the corresponding business operation. For example, the operational service "SRTS01.03. Initiation of the vehicle registration process" (**Fig. 3**) defines the following automated functions:

– Verify the originating message for authenticity

– Go to the transfer of documents for registration/re-registration of the vehicle

– Go to pay state fees

– Go to the choice of traffic police appointments

– Complete the preparation for registration / re-registration of the vehicle.

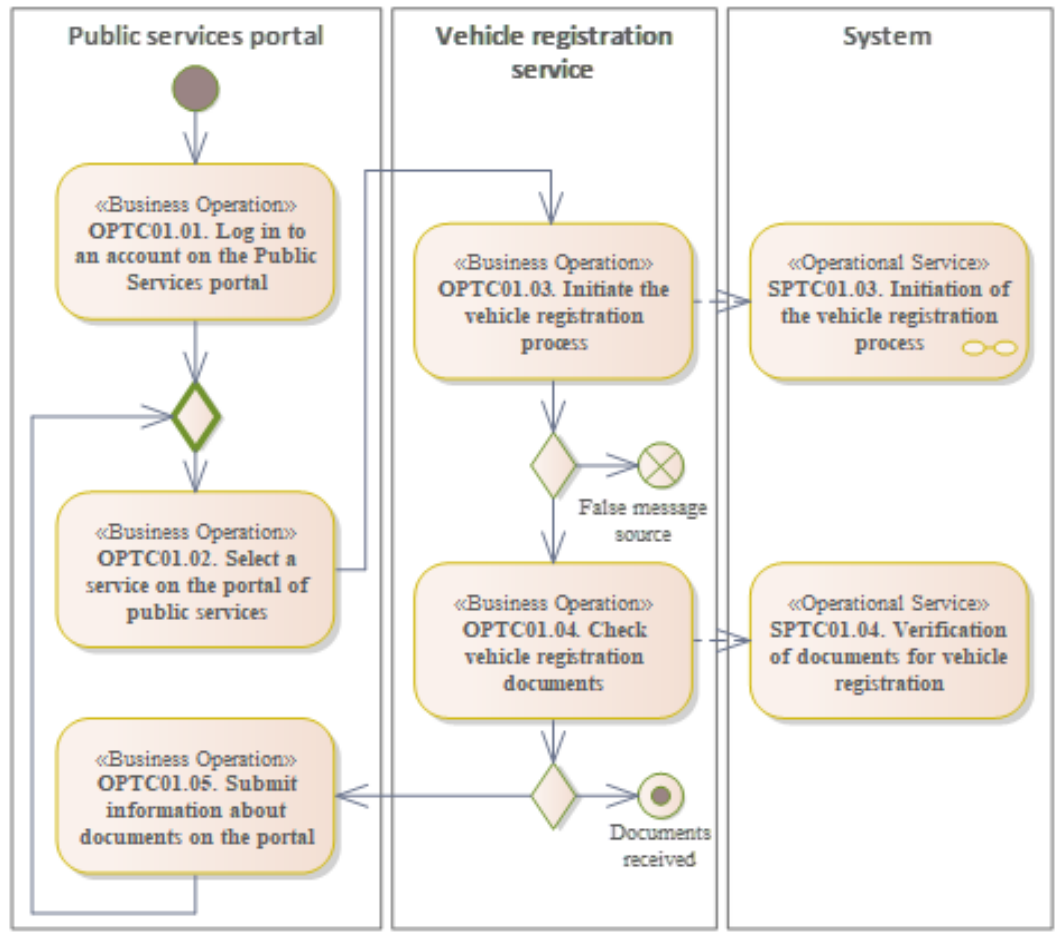

**Fig. 3.** Decomposition of business function F01

## 4. FUNCTIONAL ARCHITECTURE

The functionality specified by the operational service is implemented by a multitude of interconnected system dialogs between system agents. An operational service script is a model of related dialogs that implement the automated functions that define a service. The service scenario diagram is formed as a decomposition of the operational service (**Fig. 4**).

A system dialogue is understood as any act of interaction between agents that causes a change in the state of the IS by launching the corresponding software components. Thus, dialogue is understood in a broad sense - it is not only the interaction of the user with the system, but also the exchange of messages between any IS objects. Dialogues are described by (a) a form (if it is a user dialogue with the system); (b) functions of the IS presentation layer, i.e. view functions that provide access to the implementation of automated functions; (c) various constraints (preconditions, postconditions, error handling, etc.), which are also considered as view functions (**Fig. 5**).

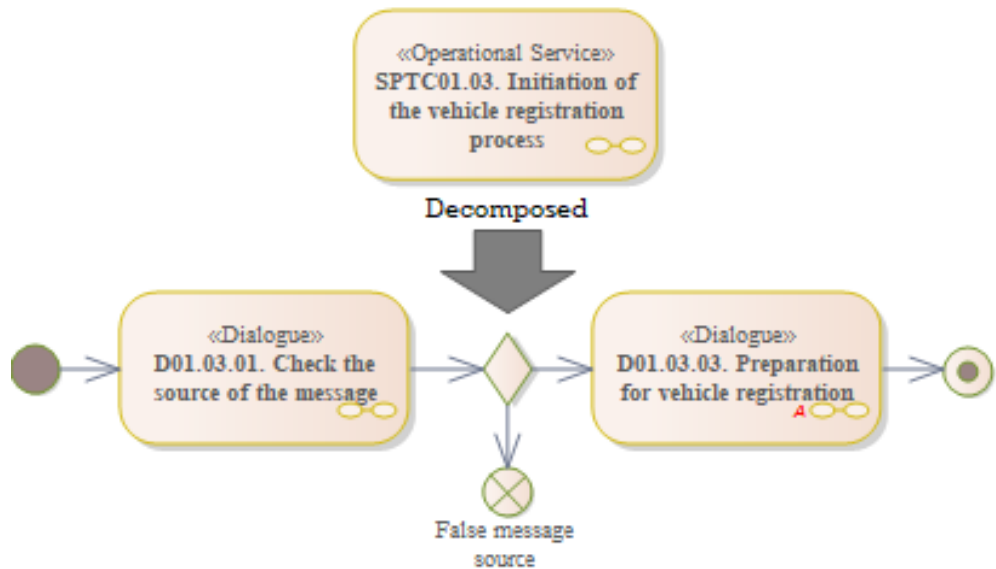

**Fig. 4**. Operational service script

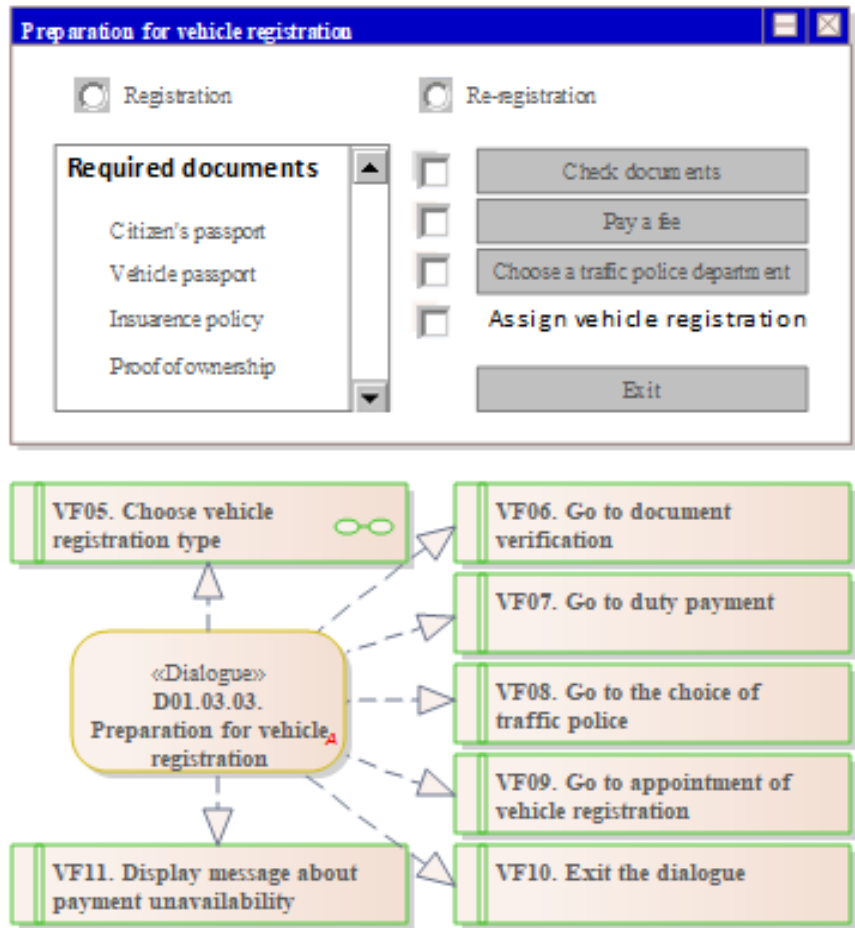

**Fig. 5.** Dialog definition (dialog form and view functions)

The report on the dialog D01.03.03, obtained on the basis of the presented diagram using the Enterprise Architect report generator [13], is presented below and includes a description of the dialogue form and view functions (Table 1).

Dialog D01.03.03. Preparation for vehicle registration

The screen shows a form on which are located:

- alternative switches "Registration" and "Re-registration" to select the type of procedure required by the User;
- a list of required documents, in which checked documents transferred from the public services portal are marked;
- button "Check documents" with a flag marking the completed check;
- button "Pay duty" with a flag marking the payment of the fee;
- button "Select State traffic inspectorate" with a flag of the selection made;
- checkbox "Assign registration", finalizing the preparation for registration of the vehicle;
- "Exit" button to exit the preparation for vehicle registration mode.

Table 1. View functions of the dialog D01.03.02. Preparation for vehicle registration

| View function | Function definition |
|---|---|
| VF05. Choose vehicle registration type | The user selects one of the alternative switches ("Registration" or "Re-registration") depending on the type of procedure he needs. |
| VF06. Go to document verification | To check the documents transferred from the public services portal, the User clicks the "Check documents" button. |
| | Control is transferred to the dialog "D01.04.01. Submit documents for registration of the vehicle". Upon completion of the verification of the completeness of information about the documents received from the public services portal, the documents that have passed the verification are marked on the dialog form and a checkbox is ticked indicating that the documents have been verified. |
| | When you call the dialog again, in case of adding missing documents on the portal of state services, the button "Check documents" can be pressed again. |
| VF07. Go to duty payment | To pay the fee for registration/re-registration of the vehicle, the User clicks the "Pay the fee" button. |
| | Control is transferred to the dialog "D03.01.01. Generate and transfer an invoice for payment of state duty". Upon completion of the payment of the fee, the checkbox is automatically ticked, signaling the payment of the fee. |
| | Upon payment of the state duty, the button "Pay the duty" is blocked and pressing the button again becomes impossible. |
| VF08. Go to the choice of traffic police | To select the traffic police department and the time of the visit, or to cancel the visit, the User clicks the "Choose traffic police department" button. |
| | Control is transferred to the dialog "D02.01.01. Choose traffic police department or cancel the visit". Upon completion of the selection of the traffic police department, a checkbox is automatically ticked, signaling the choice of the time of visit. If the visit is canceled, the checkbox is cleared. |

| View function | Function definition |
|---|---|
| VF09. Go to appointment of vehicle registration | The checkbox "Assign vehicle registration" is available for setting or unchecking if the checkboxes setting next to the buttons<br>Check documents<br>Pay the fee<br>Choose the traffic police department.<br>To complete preparation for registration of the vehicle, the user selects the "Assign vehicle registration" checkbox.<br>To cancel registration, the User unchecks this checkbox. At the same time, the checkbox is unchecked at the button "Choose the traffic police department", and the previously appointed visit time is released. |
| VF10. Exit the dialogue | At any time, the User can end the dialog by clicking the "Exit" button. At the same time, the current state of the service of registration / re-registration of the vehicle is preserved, and control is transferred to the dialog of the business operation "OPTC01.02. Select a service on the portal of public services". |
| VF11. Display message about payment unavailability | If an error occurs when calling the payment service of the public services portal, a message is displayed:<br>*C002. Payment of the state fee is not available now. Try again later*<br>The system remains on the same screen that was displayed to the User at the time of the request. |

## 5. COMPONENT ARCHITECTURE

The view functions resulting from the design of dialogs do not take into account the component architecture of the system. These functions concern only the presentation level of the IS. Representation of the IS architecture in the form of interacting functional components (subsystems, web services, external IS) makes it possible to find out how the view functions are implemented, to determine the internal functions of the application logic and data management. The component architecture establishes the composition and interaction of the functional components of the IS, determines the software modules and their distribution according to the functional components, details the functional requirements for the IS.

The main elements of the component architecture model are:
– functional components, determined by the choice of an architectural pattern (or a composition of architectural patterns) and the external environment of the system;
– software modules, which are the structural parts of the functional components of the IS and are determined by the decomposition of the view function.

The design of component architecture begins with the definition of the component structure of the IS. It represents the system as the set functional components (subsystems and external systems). The modular composition of the functional components of the system is determined by the decomposition of the elements of the functional architecture - the view functions of IS, performed on the selected component structure. The decomposition of the view function may include previously defined modules for reuse, i. e. there is a many-to-many relationship between view functions and system modules.

In ACM, to determine the modular composition of components of the IS, Sequence Diagrams are used, which are designed for the view functions of each dialog described at the level of functional architecture. Functional components are used as lifelines in diagrams. Thus, design allows you to define a complete set of software modules for all functional components.

The following functional components are defined for the vehicle registration service:
– a portal of public services external to the service;
– information security subsystem;
– message reception / transmission subsystem;
– application server subsystem
– data storage subsystem.

Definition of subsystems modules for the view function "VF05. Choose vehicle registration type" is performed by decomposition on the component structure (**Fig. 6**).

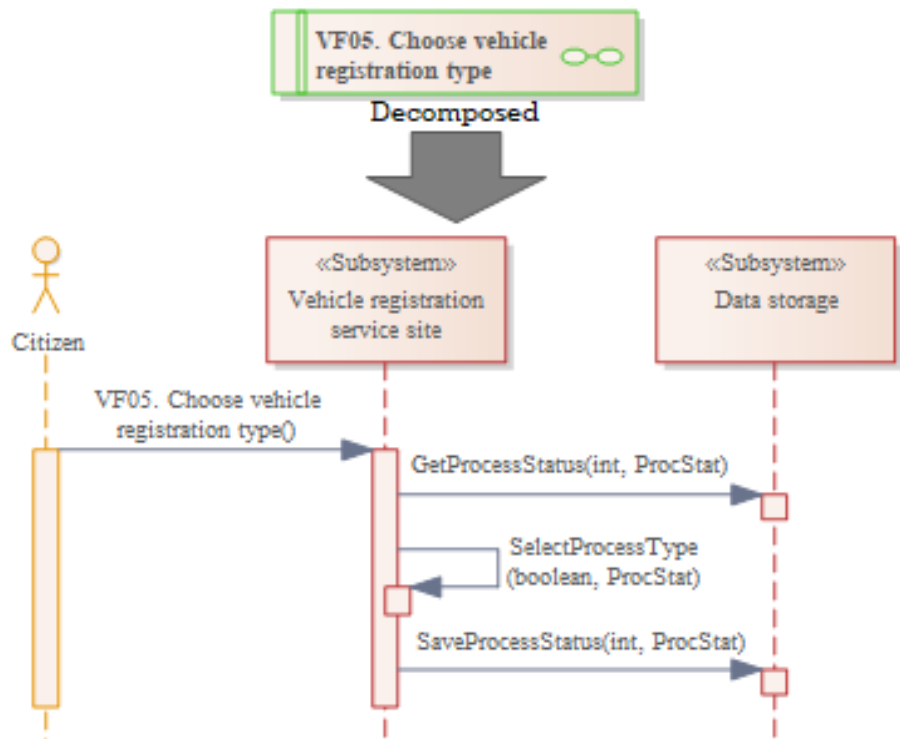

**Fig. 6**. Definition of IS component modules

Table 2, generated from the presented sequence diagram, provides a report on the relationship of software modules with view functions and their distribution among the elements of the component architecture.

**Table 2.** Defining the modules of the view function VF05. Choose vehicle registration type

| Module | Parameters | Component | Comments |
|---|---|---|---|
| GetProcessStatus | UserID, ProcessCondition | Data storage | For the User, the current state of preparation for registration / re-registration of the vehicle is determined:<br>– previously selected alternative switch Registration/Re-registration;<br>– verified documents;<br>– payment of state duty;<br>– appointment with the traffic police department;<br>– whether vehicle registration has been assigned. |
| SelectProcessType | OptionID, ProcessStatus | Vehicle registration service site | The user sets one of the alternative switches ("Registration" or "Re-registration"), depending on the type of procedure he needs.<br>When changing the switch selection, the alternate switch is reset. |
| SaveProcessStatus | UserID, ProcessStatus | Data storage | The current state of preparation for the vehicle registration is saved. |

## 6. DATA ARCHITECTURE

ACM proposes technology for data architecture design [12], but the description of this technology is outside the scope of this work. We proceed from the fact that the data architecture is somehow developed at the level of the ER model and includes all the necessary classes distributed between the elements of the component model.

The task is to determine the functional components of the data architecture (class methods) based on the formed component structure. For each module, the order of its implementation within the framework of the object-oriented IS model is determined.

Methods are defined by decomposition of system modules on a given class model. The diagram (**Fig. 7**) demonstrates the implementation of the GetProcessStatus module with the corresponding class methods.

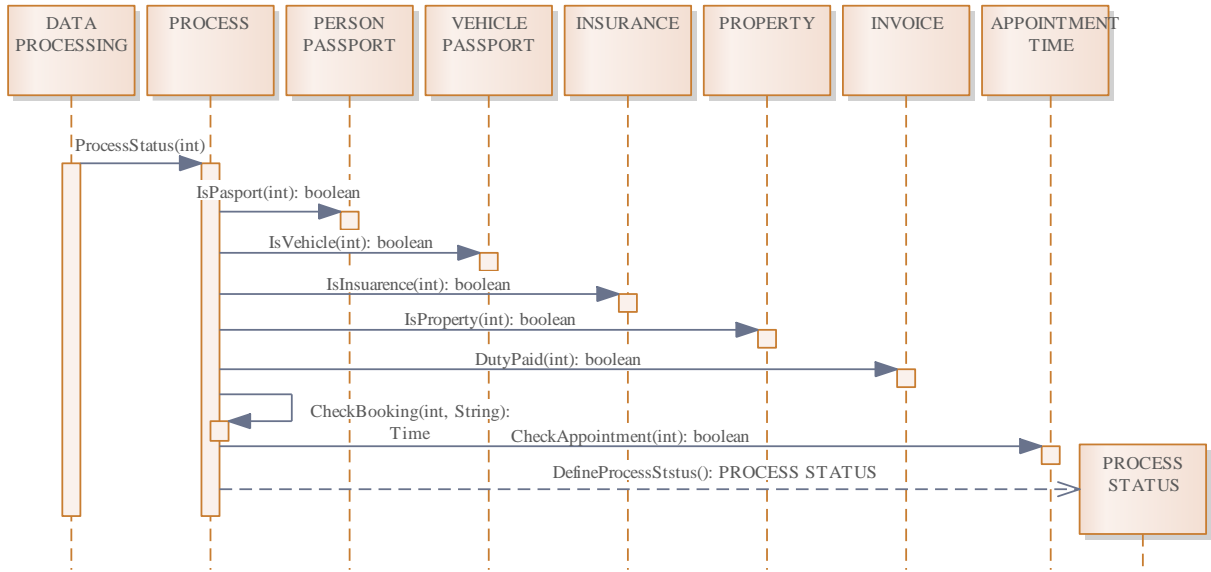

**Fig. 7.** GetProcessStatus code module class methods

The description of the methods of the classes presented in the diagram is given in table 3.

**Table 3.** Description of the methods that implement the GetProcessStatus module

| Method | Parameters | Class | Comments |
|---|---|---|---|
| ProcessStatus | ProcessID | PROCESS | A request is initiated to determine the state of the current process. The state is determined by the following parameters: availability of documents required for vehicle registration payment of duty choice of time of registration registration appointment. |
| IsPasport | ProcessID | PERSON PASSPORT | The presence of passport data is checked (the Is_Passport attribute). |
| IsVehicle | ProcessID | VEHICLE PASSPORT | The presence of vehicle passport data is checked (Is_Vehicle attribute). |
| IsInsurance | PassportID | INSUARENCE | The existence of an insurance policy is checked (the Is_Insurance attribute). |
| IsProperty | ProcessID | PROPETY | The presence of a document certifying the ownership of the vehicle (the Is_Property attribute). |
| DutyPaid | ProcessID | INVOICE | The payment of the state duty for vehicle registration is checked (the Payment_Result attribute). |
| CheckBooking | ProcessID, TrafficPolice | PROCESS | The timing of the visit to the traffic police is checked (there is a link to the APPOINTMENT TIME entry). |

| Method | Parameters | Class | Comments |
|---|---|---|---|
| CheckAppointment | ProcessID | APPOINTMENT TIME | The appointment of the traffic police visit is checked (the Appointment_Booking attribute). |
| DefineProcessStstus | | PROCESS STATUS | A class is created that stores information about the status of the current registration process. |

## 7. CONCLUSION

Techniques considered in this paper provides a seamless transition from business architecture through the decomposition of the operational services to the functional architecture of the IS, which defines the dialogs of the system and the view functions. The decomposition of the view functions defines the software modules of the component architecture and, finally, the modules are decomposed into classes methods of the data architecture. In other words, each architectural representation is derived from the architecture of the previous level of abstraction.

With this approach, the completeness of the functional implementation of the IS is ensured, since the decomposition of the business process allows you to accurately formulate the user requirements for the automated functions. Further design of functional components at various levels of the abstract description of IS is essentially a reasonable conclusion of the necessary and sufficient functionality of the IS, which avoids errors associated with technological gaps between architectural models, insufficient or excessive functionality.

The advantages shown by the ACM in real-world design and maintenance of IS include ensuring transparency of the compliance of the design result with the requirements set by the stakeholders by simplifying the validation and verification processes. The presence of tracing between the elements of architectural models allows for the rapid localization of necessary changes and improvements for the release of new versions of IS. Management of the development process, the relationship of architectural description with artifacts, taking into account the possibility of generating design and operational documents based on architectural models [11], significantly reduces design time and eases the process of change implementation to IS.

**Acknowledgments**

The author expresses sincere gratitude to Professor Boris Pozin, who provided invaluable assistance in the preparation of this article.